\documentstyle[prb,aps,epsf,psfig]{revtex}
\begin{document}
\title{
Charge ordering and antiferromagnetic exchange in \\
layered molecular crystals of the $\theta$ type
}
\draft
\author{
 Ross H. McKenzie\cite{email0} and J. Merino\cite{email1}}
\address{Department of Physics, University of Queensland,
 Brisbane 4072, Australia}
\author{J. B. Marston}
\address{Department of Physics, Brown University, Providence, RI-02912-1843}

\author{ O. P. Sushkov}
\address{School of Physics, University of New
South Wales, Sydney 2052, Australia}
\date{\today}
\maketitle
\widetext

\begin{abstract}
We consider the electronic properties of
layered molecular crystals of the type $\theta$-D$_2$A
where A is an anion and D is a donor molecule
such as BEDT-TTF 
[where BEDT-TTF is bis-(ethylenedithia-tetrathiafulvalene)]
which is arranged in the $\theta$ type pattern within the
layers. We argue that the simplest strongly correlated
electron model that can describe the rich phase
diagram of these materials is
the extended Hubbard model on the square lattice at
a quarter filling. In the limit where the Coulomb
repulsion on a single site is large, the nearest-neighbour
Coulomb repulsion $V$ plays a crucial role.
When $V$ is much larger than the intermolecular hopping integral
$t$ the ground state is an insulator with charge ordering.
In this phase antiferromagnetism arises due to a novel
fourth-order superexchange process around a
plaquette on the square lattice. We argue that
the charge ordered phase is destroyed below a critical
non-zero value $V$, of the order of $t$.
Slave boson theory is used to explicitly demonstrate this 
for the SU(N) generalisation of the model, in the large N limit.
We also discuss the relevance of the model to the all-organic
family $\beta''$-(BEDT-TTF)$_2$SF$_5$YSO$_3$
where Y = CH$_2$CF$_2$, CH$_2$, CHF.
\\
\\
\end{abstract}

\pacs{PACS numbers: 71.27.+a, 71.10.Fd, 74.70.Kn, 71.45.Lr}

\section{Introduction}

Layered organic molecular crystals
based on  the BEDT-TTF molecule [where BEDT-TTF is 
bis-(ethylenedithia-tetrathiafulvalene)]\cite{williams,ishiguro,wosnitza}
 are rich examples of strongly correlated electron
systems in low dimensions.
The $\kappa$-(BEDT-TTF)$_2$X
family has many similarities
to the cuprate superconductors, such as the
proximity of superconductivity to an antiferromagnetic
Mott-Hubbard insulating phase.\cite{kino,kanoda,mck}
It has recently been argued that the
simplest strongly correlated electron model
that can describe 
this family is a Hubbard model on an anisotropic triangular
lattice at half-filling.\cite{mck} This model should also describe
the $\beta$-(BEDT-TTF)$_2$X family.\cite{merino,okuno}
As the anion X or pressure is varied
the $\theta$-(BEDT-TTF)$_2$X  family
exhibits metallic, superconducting, insulating,
 antiferromagnetic, charge ordered,
and spin gapped phases.\cite{mori2,co}
The family $\theta$-(BETS)$_2$X [where BETS is
bis-(ethylenedithio-tetraselenafulvalene)]\cite{sulphur}
is also found to exhibit a
metal-insulator transition with a transition temperature
which varies with the anion.\cite{mori4}
 The recently synthesized family $\beta''$-(BEDT-TTF)$_2$SF$_5$YSO$_3$
where Y = CH$_2$CF$_2$, CH$_2$, CHF
has attracted considerable interest from chemists because
the anion is purely organic and 
Y = CH$_2$CF$_2$
is the first purely organic superconductor.\cite{wang}
Insulating, charge ordered and spin gapped phases
are observed for different anions.\cite{ward}     

Given the complexity of the details of the 
chemistry, crystal structure  and band structures
of these materials it is important to define
the simplest possible many-body Hamiltonian
that can capture the essential physics.\cite{anderson}
This is similar in spirit to the way one
studies the Hubbard and $t-J$ models on a square
lattice in order to understand the cuprate superconductors.
Several previous studies\cite{mori3,mori4} of the metal-insulator transition
and the magnetic properties of the insulating phase\cite{yamaura}
of the $\theta$ family have been interpreted in terms of the Hubbard model.
However, this is inadequate because, at quarter filling, the
Hubbard model is expected to be metallic. 
Following Seo\cite{seo}, we argue that the nearest neighbour Coulomb repulsion
$V$ plays a crucial role in these materials.
This has been emphasized before
for other organics.\cite{Mila,co,enoki,mazumdar}

Specifically, the simplest possible strongly
correlated electron model that can describe
the competition between all of the above phases is
the extended Hubbard model (or a $t-V$ model with no
double occupancy) at quarter-filling on a square lattice.
We show that in the charge-ordered
insulating phase antiferromagnetic interactions
arise due to a novel fourth-order superexchange around
a plaquette on the square lattice.
We then consider the SU(N) generalization of the
$t-V$ model and perform a slave boson study which
becomes exact in the limit of large N.
We find there is a critical value of the ratio $V/t$,
above which the metallic Fermi liquid phase undergoes
an instability to a charge ordered state.

\subsection{ Review of experimental properties of $\theta$-(BEDT-TTF)$_2$X}

Properties of 
the $\theta$-(BEDT-TTF)$_2$X  family
have recently been reviewed by Mori {\it et al.}\cite{mori3,mori}
and Seo\cite{seo}.
The arrangement of the BEDT-TTF or BETS molecules within
a layer of the $\theta$ structure is shown in Fig. \ref{fig1}.
For most anions, X, the materials undergo a metal-insulator
transition at a temperature that decreases with
increasing band-width; the latter is directly
correlated with the angle between the  molecules
within the layers\cite{mori3} (see Fig. \ref{fig2}). 
The temperature at which the metal-insulator transition
occurs generally increases with increasing pressure.\cite{mori4}

$\theta$-(BDT-TTP)$_2$Cu(NCS)$_2$ undergoes
a metal-insulator transition at 250 K.\cite{ouyang}
At low temperatures the charge gap deduced from
the activation energy of the conductivity is about 100 meV.
 Raman-active vibrational
modes (associated with the stretching of
carbon-carbon double bonds) of the BPT-TTP molecules
are sensitive to the charge on the molecule.
In the insulating phase these modes split,
consistent with charge ordering.\cite{ouyang}
Evidence for short-range charge ordering along the
c-axis  direction (the vertical direction in Fig. \ref{fig1})
 was found
in the insulating phase of X=CsCo(SCN)$_4$ by x-ray scattering.\cite{watanabe}
For X=CsZn(SCN)$_4$, the principal axes of the g-tensor
associated with electron spin resonance undergo a rotation
at 20 K; this has been interpreted as a change in
the electronic state.\cite{nakamura2}
For X=RbZn(SCN)$_4$ 
there is a metal-insulator transition
at 190 K; there is then a dimerization in
the c-direction.\cite{mori3}
 The magnetic susceptibility shows no features at
this transition and between
50 and 190 K has been fit to that for
a two-dimensional antiferromagnetic              Heisenberg
model with exchange $J$  = 100 K.
Below 50 K the susceptibility is consistent with
a spin gap of about 45 K.
There is evidence from nuclear magnetic resonance
for charge ordering below 190 K and of a spin gap below 10 K.\cite{Miyagawa}
For X= Cu$_2$(CN)[N(CN)$_2$]$_2$
there is a metal-insulator transition
at 220 K; the charge gap at low temperatures
is about 200 meV.\cite{komatsu}
 The magnetic susceptibility shows no features at
this transition and between
33 and 220 K has been fit to that for
a two-dimensional antiferromagnetic              Heisenberg
model with $J$ = 48 K.
There is no sign of Neel order but below 30 K
the susceptibility decreases rapidly, suggesting 
formation of a spin gap.
The only member of the $\theta$-(BEDT-TTF)$_2$X
 family that is superconducting
is X=I$_3$ which has a transition temperature of 3.6  K.
The Fermi surface of the metallic phase has been mapped out
using angular-dependent magnetoresistance and magnetic
 oscillations.\cite{tamura}
Several of the 
$\theta$-(BETS)$_2$X  family undergo a metal-insulator
transition  and
several are metallic down to 4 K  (see Table \ref{table1}). 

$\theta$-(C$_1$TET-TTF)$_2$Br [where C$_1$TET-TTF is
bis(methylthio)ethylenedithio-tetrathiofulvalene]
is an insulator with a charge gap of 600 meV.\cite{yamaura}
 The magnetic susceptibility between
8 and 290 K has been fit to that for
a two-dimensional antiferromagnetic  Heisenberg
model with $J$ = 6 K.
Below 3 K the susceptibility depends on
the field direction, suggesting the formation
of Neel order.

As emphasized by Mori\cite{Mori}, and illustrated
schematically in Figure \ref{fig2}, Table \ref{table1} shows
the general trend that as the band width (which
is roughly proportional to $t_p$) increases the
transition of the metal-insulator transition decreases.

\section{The extended Hubbard model}

The arrangement of the BEDT-TTF molecules within a 
layer of $\theta$-(BEDT-TTF)$_2$X
is shown schematically in Figure 1.
Values of the intermolecular hopping integrals,
calculated using the  H\"uckel approximation are given in Table \ref{table1}.
If there is complete charge transfer of one electron
onto each anion X 
there is an average of half a hole per molecule and
so the electronic bands will be quarter filled.

Band structure calculations\cite{ward,yamaura,komatsu,Mori,kobayashi}
  predict that
all these materials are metallic. Hence,
the different phases must be due to strong
electronic interactions.
The Hubbard interaction $U$ describing the
Coulomb repulsion between two electrons
on the same BEDT-TTF molecule has been estimated
from quantum chemistry
calculations to be about 4 eV.\cite{painelli,castet}  
Thus $U$ is much larger than the band width associated
with the hopping integrals given in Table 1,
confirming that these are strongly correlated materials.
The  Hubbard model on the square lattice
at quarter filling is expected to be always metallic 
and so one must consider longer range
Coulomb interactions to explain the existence
of insulating, charge ordered and antiferromagnetic phases.
The nearest neighbour Coulomb repulsion
between various arrangements of pairs of BEDT-TTF molecules
has also been estimated from
quantum chemistry calculations\cite{castet,painelli,moriV}
and is generally  found to have a value of about 2-3 eV.
It is approximately given by Coulomb's law $V \simeq 14 \ {\rm eV}/R$,
where $R$ is in $\AA$. Mori calculated
$V$ as a function of the angle $\theta$;
variations of about ten per cent occur in
the range (98 - 130 degrees) relevant to
the $\theta$-type materials.  These calculations involve isolated
pairs of molecules and so one expects
that the values of $U$ and $V$ in a molecular crystal to
be less than this due to screening.
Hubbard has discussed this for the case of TTF-TCNQ
arguing that $U$ and $V$ are both decreased by a factor
of about two.\cite{hubbard}
Actually in Section \ref{afm}, from experimentally determined
charge gaps, we estimate values of $V$ of the order
of a few hundred meV.
In materials consisting of dimers of BEDT-TTF molecules
the difference $U-V$ can be estimated from the charge
transfer excitation seen in optical absorption measurements.
For (BEDT-TTF)HgBr$_3$ it is estimated to be 0.7 eV.\cite{tajima}
Thus we are led to the extended Hubbard model on
the anisotropic triangular lattice at half filling.

Table \ref{table1} shows that for many of the $\theta$ 
materials, $t_c \ll t_p$ and so, as 
a first approximation, we neglect the diagonal hopping $t_c$.
This means we are left with a square lattice model.
In Section \ref{slave} we will show that this diagonal term
has little effect on the metal-insulator transition.
The Hamiltonian is
\begin{equation}
H = t \sum_{<ij>,\sigma} 
(c^\dagger_{i \sigma} c_{j \sigma} + 
c^\dagger_{j \sigma} c_{i \sigma})
+ U \sum_{i} n_{i\uparrow} n_{i\downarrow}
+ V \sum_{<ij>} n_i n_j - \mu \sum_{i \sigma} n_{i \sigma}
\label{ham}
\end{equation}
where $U$ is the Coulomb repulsion between two electrons on
the same site, $V$ is the nearest
neighbour Coulomb repulsion, 
$<ij>$ denotes nearest neighbours,  and $\mu$ is the chemical potential.

If we consider the large $U$ limit and so preclude
doubly occupied sites,
the Hamiltonian then reduces to
\begin{equation}
H = t \sum_{<ij>,\sigma}  
P (c^\dagger_{i \sigma} c_{j \sigma} + 
c^\dagger_{j \sigma} c_{i \sigma}) P
+ V \sum_{<ij>} n_i n_j - \mu \sum_{i \sigma} n_{i \sigma}
\label{ham2}
\end{equation}
where $P$ projects out the doubly occupied states.
We refer to this as the $t-V$ model.

For large $V/t$ the ground state is an
insulator with charge ordering (Section \ref{afm}).
We expect that for small $V/t$ a metallic 
phase exists because the quarter-filled
Fermi surface is poorly nested.
Hence, as $V/t$ decreases the charge ordering should be destroyed at a non-zero 
value of $V/t$.  We are not aware of any previous study of
this model on the square lattice.
Mazumdar, Clay, and Campbell\cite{mazumdar} have studied
coupled chains of the  extended Hubbard model at quarter filling.
Most of their numerical results
are for $U=6t_\parallel$ and $V_\parallel=t_\parallel$
(where the $\parallel$ refers to the chain direction) so that
$V_\parallel$ is smaller than the critical value necessary
to form the charge ordered state considered here.
They find that when interactions with phonons are included
there is a tendency to formation of a bond-order wave.
 Henley and Zhang recently
considered a similar model involving spinless fermions
on the square lattice.\cite{henley}
Mila considered the extended Hubbard model on the 
square lattice at a quarter filling with finite $U$ and
infinite $V$.\cite{Mila2}
In Section \ref{slave} we consider the
SU(N) generalization of the $t-V$ model and
show that in the large N limit, slave boson theory
implies that there is a critical value of $V/t$ above
which the metallic phase becomes unstable
to formation of charge ordering.

 We now briefly review previous
work on the extended Hubbard model (in the large $U$ limit)
at quarter filling on different lattices.
Numerical calculations show that a
single chain undergoes a transition from a Luttinger liquid
to a charge-density wave insulator 
at $V=2t$.\cite{nakamura}
The one-dimensional $t-V$ model can be solved exactly via
the Bethe ansatz\cite{schlottmann}. It is equivalent
to two decoupled XXZ spin chains and so will undergo
a metal-isulator transition at $V=2t$.
We note that the ring exchange process
responsible for antiferromagnetic interactions
discussed above will be absent in a single chain.
Vojta, H\"ubsch, and Noack\cite{votja} recently studied the
model (\ref{ham}) on a ladder using the density matrix
renormalisation group.
They find that the charge ordered phase is
destroyed for $ V < 2.5 t$ but claim that there
will be a charge gap for all $V/t$.
The model in infinite dimensions was studied by 
Pietig, Bulla, and Blawid using
dynamical mean-field theory.\cite{bulla}
At low temperatures they found that for $U=2 t$,  
charge ordering occured for $V > 0.5 t$.

\section{Antiferromagnetic exchange in the charge-ordered phase}
\label{afm}

For large $V/t$ there will be charge ordering and
the ground state will be an insulator with a charge
gap of magnitude $3V$. There will be two possible
ground states with the checker-board covering of the lattice
 (Fig. \ref{fig3}).
This defines a new square lattice rotated by
45 degrees with respect to the original square lattice.
It should be stressed that these ground states are distinct
from the commensurate charge density waves, associated with
a lattice distortion, and seen in some organics.
To zero-th order in $t/V$ all possible spin states
will be degenerate.
 We consider a single placquette (Fig. \ref{fig3}) containing two spins.
To second order in $t/V$ both the singlet and
triplet states have their energy lowered
by $-4 t^2/3V $.
 The degeneracy
of the singlet and triplet states is only broken to fourth order in $t/V$. 
We show below that this results in an effective antiferromagnetic exchange
interaction 
\begin{equation}
J = { 4 t^4 \over 9 V^3}
\label{jeff}
\end{equation}
which acts along the diagonals of the original 
square lattice. 
 Thus the spin degrees of freedom are described by
a spin-${1 \over 2}$ Heisenberg model on a square lattice.
The Hamiltonian is
\begin{equation}
\label{ham3}
    H = J \sum_{\rm \langle {\bf ij} \rangle} \bbox{S}_{\bf i} \cdot
 \bbox{S}_{\bf j}
\end{equation}
where $\bbox{S}_{\bf i}$ denotes a spin operator
on site ${\bf i}$, 
and    the sum ${\rm \langle {\bf{ij}} \rangle }$
runs over pairs of  next-nearest neighbor lattice sites
in the original square lattice.

We now calculate
the singlet-triplet splitting from fourth-order
perturbation theory.
If $| \Psi_0 >$ is an eigenstate of $H_0 = V \sum_{<ij>} n_i n_j $ then
a perturbation $H_1 = H - H_0$, which has no effect
to third order, shifts the energy by 
\begin{equation}
\Delta E_0^{(4)} =  
\sum_{\{m,n,p\} \neq 0}
{< \Psi_0 | H_1 | \Psi_m > < \Psi_m | H_1 | \Psi_n > < \Psi_n | H_1 | \Psi_p >
< \Psi_p | H_1 | \Psi_0 > \over
 (E_0^{(0)} -E_m) (E_0^{(0)} -E_n) (E_0^{(0)} - E_p )
}
\label{pert4}
\end{equation}
where the intermediate states labelled
 by $\{m,n,p\} \neq 0 $ are not degenerate
with $| \Psi_0 >$.
The following process
involving exchange of electrons around a placquette
will contribute to a shift in the energy of the ground state.
For the triplet states it can be represented as
\begin{equation}
\left| \begin{array}{cc}
\uparrow &  o \\
o & \uparrow \\ \end{array}\right>
\  \  \rightarrow \ \
\left| \begin{array}{cc}
o & \uparrow  \\
o & \uparrow \\ \end{array}\right>
\  \  \rightarrow \ \
\left| \begin{array}{cc}
o & \uparrow  \\
 \uparrow & o \\ \end{array}\right>
\  \  \rightarrow \ \
\left| \begin{array}{cc}
 \uparrow  & \uparrow  \\
 o  & o \\ \end{array}\right>
\  \  \rightarrow \ \ 
\left| \begin{array}{cc}
 \uparrow  & o \\
 o  & \uparrow   \\ \end{array}\right>
\end{equation}
The first, second, and fourth
matrix elements are $t$ and the third is $-t$.
The intermediate states have energy $3V$, $4V$, and $3V$, respectively.
(Note one needs to take into account the interaction with the
neighbours not shown in the above representation).
There are eight distinct ways of doing this exchange:
at the first step there is four choices
and at the third step there are two choices.
Thus, the expression (\ref{pert4}) implies that
the triplet is increased in energy by $ 2t^4/9 V^3$.

A similar process for the singlet is 
\begin{equation}
\left| \begin{array}{cc}
\uparrow &  o \\
o & \downarrow \\ \end{array}\right>
\  \  \rightarrow \ \
\left| \begin{array}{cc}
o & \uparrow  \\
o & \downarrow \\ \end{array}\right>
\  \  \rightarrow \ \
\left| \begin{array}{cc}
o & \uparrow  \\
 \downarrow & o \\ \end{array}\right>
\  \  \rightarrow \ \
\left| \begin{array}{cc}
 \downarrow  & \uparrow  \\
 o  & o \\ \end{array}\right>
\  \  \rightarrow \ \
\left| \begin{array}{cc}
 \downarrow  & o \\
 o  & \uparrow   \\ \end{array}\right>
\end{equation}
Thus, for the singlet this fourth-order process brings
one back to the singlet wave function with a sign change and
so the energy shift is opposite to that of the triplet.
 Hence, we arrive at (\ref{jeff}) for the difference
in energy between the singlet and triplet.

It should be pointed out that there are also
fourth order processes of the form
$|0> \rightarrow |n> \rightarrow |0> \rightarrow |p> \rightarrow |0>$
which will produce a decrease in the ground state energy.
However, because they produce the same shift for
the singlet and triplet states we neglect them here.
Our value of $J$ is consistent with a recent calculation\cite{votja}
of the effective exchange
interaction in the charge ordered phase
of the extended Hubbard model on a ladder when that
result is rescaled to allow for
different excitation energies of the intermediate
states.
For the ladder, the energies of the intermediate states
are all $2V$.
Thus the ladder exchange is larger than for the
square lattice by a factor of 9/2.

We now consider whether this possible explanation
for the origin of antiferromagnetism in the $\theta$ type 
materials is quantitatively reasonable.
Roughly, it predicts that the value of $J$ will be some fraction 
of $t$, typical values\cite{caution} from Table \ref{table1}
are of the order of 500-1000 K for the materials
with insulating ground states.
$\theta$-(C$_1$TET-TTF)$_2$Br 
is an insulator with a charge gap of 600 meV
and a value of $t$\cite{caution} of about 60 meV.\cite{yamaura}
Since the charge gap is $3V$ for $t \ll V$ this gives
$V = $ 200 meV. Using 
$ J =4t^4/9 V^3$ gives $J \sim $ 4 K (check)
which is in reasonable agreement with the observed value
of 6 K.
We do not make quantitative comparisons of equation (\ref{jeff})
because they are not so clearly in the regime
$t \ll V$, required for its validity.
For example, for $\theta$-(BEDT-TTF)$_2$Cu$_2$(CN)[N(CN)$_2$]$_2$
H\"uckel calculations give $ t \sim$ 80 meV and
the measured charge gap\cite{komatsu} is about 200 meV.
This is inconsistent with the fact that the charge gap
would be $3V$ if $t \ll V$.

\section{ Slave boson theory for the SU(N) version
of the extended Hubbard model with $U \rightarrow \infty$}
\label{slave}

 We consider the SU(N) generalization
of the Hamiltonian
(\ref{ham2}) for which the spin index, $\sigma$, runs from 1 to $N$,
and consider $1/N$ as a small expansion parameter assuming that
$N$ is large.
 Slave boson fields are introduced to allow  treatment
of the no double occupancy
constraint required by the $U \rightarrow \infty$ limit.
The effective action for the slave boson fields can be
expanded in powers of $1/N$.
The mean-field solution corresponds to the Gutzwiller approximation
and is exact in the $N \rightarrow \infty$ limit.\cite{Read1,Kotliar0}
This approach has been used
to study other strongly correlated models such as the
Kondo model for magnetic impurities\cite{Read0,Read1} in metals,
 the Hubbard model\cite{Kotliar0}, the Hubbard-Holstein model\cite{becca}
and the Anderson\cite{Millis}
 and Kondo lattices.\cite{Auerbach}
 It has also been used to analyze
the phase diagram of two-dimensional
t-J model\cite{Kotliar1,Grilli,Sachdev}, and extended Hubbard model at
half-filling.\cite{caprara}
It is convenient to describe the projected Hilbert space
associated with the Hamiltonian (\ref{ham2}), using the slave boson
representation.\cite{Barnes,Coleman,Read0} The electron
creation operator is replaced by:
$c^{\dagger}_{i\sigma}=f^{\dagger}_{i \sigma} b_i$, where the
spinless charged boson operator, $b_i$, is introduced to keep
track of the empty sites, and $f^{\dagger}_{i \sigma}$ is a fermion
operator carrying spin. In order to  preserve
the anticommutation relation for the electrons
the new operators must satisfy the local constraint
\begin{equation}
f^{\dagger}_{i\sigma}  f_{i\sigma}+ b^{\dagger}_i b_i=N/2.
\label{constraint}
\end{equation}
For $N=2$ it reduces to the condition that either an
electron or a boson can occupy each lattice site at all times.

Following Kotliar and Liu \cite{Kotliar1},
we write the partition function in the coherent state path
integral representation:
\begin{equation}
Z= \int D b^{\dagger}D b D f^{\dagger} D f D \lambda
\exp\left(-\int_{0}^{\beta} L(\tau) d \tau \right)
\label{z}
\end{equation}
where the Lagrangian at imaginary time $\tau$ is given by
\begin{eqnarray}
L(\tau) &=& \sum_{i}
 f^{\dagger}_{i\sigma}( \partial_\tau - \mu) f_{i\sigma} +
b^{\dagger}_i\partial b_i - {1\over N} \sum_{ij} T_{ij}
(f^{\dagger}_{i\sigma} f_{j \sigma} b^{\dagger}_j b_i + h.c. )
\nonumber \\
 &+&{1 \over N} \sum_{ij} V_{ij}
f^{\dagger}_{i\sigma} f_{i\sigma} f^{\dagger}_{j\sigma'}
f_{j\sigma'} + \sum_{i} i \lambda_i (f^{\dagger}_{i\sigma}
f_{i\sigma} +
 b^{\dagger}_i b_i - N/2),
\nonumber \\
\label{lagran}
\end{eqnarray}
$\beta=1/(k_B T)$ at temperature $T$, and we
have used the fact that 
$c^{\dagger}_{i\sigma}  c_{i\sigma}=f^{\dagger}_{i\sigma}  f_{i\sigma}$.
 $\lambda_i$ is a static Lagrange multiplier enforcing
the constraint (\ref{constraint}).
A sum from 1 to N is assumed whenever a repeated
$\sigma$ index appears in the equations.
The $1/N$ factors are introduced so that the
Lagrangian is proportional to $N$. $T_{ij}=t$,
if $i$ and $j$ are nearest neighbours,
$T_{ij}=t'$ if $i$ and $j$ are next-nearest neighbours sitting along
{\it one} of the diagonals of the square lattice, and $T_{ij}=0$ otherwise.
$V_{ij}=V$ if $i$ and $j$ are nearest neighbours and is zero 
otherwise.
$\mu$ is the chemical potential, which is fixed to give the
average number of electrons per site,
$n= < f^{\dagger}_{i \sigma} f_{i\sigma} >$.
The conservation of the charge, $ q=N/2$,
is a consequence of a local $U(1)$ symmetry because
under the local gauge transformation:
$b_i\rightarrow b_i e^{i \theta_i(\tau)}$,
$f_{i\sigma} \rightarrow f_{i \sigma} e^{i \theta_i(\tau)}$, and
$\lambda_i \rightarrow \lambda_i - \partial_{\tau} \theta_i(\tau)$,
$L(\tau)$ remains the same. To avoid possible infrared divergences
it is convenient to choose the radial gauge where the boson
amplitude becomes a real number, $r_i=|b_i|$, and $\lambda_i$ becomes a
dynamical field: $\lambda_i(\tau)$.
We introduce these fields in expression (\ref{lagran}),
and we use the relation $f^{\dagger}_{i\sigma} f_{i\sigma} =
N/2 - b^{\dagger}_i b_i $ to replace one
pair of fermion operators in the $V$ term so that we are
left with a quadratic Lagrangian in the fermionic
Grassmann variables.
After integrating out the fermions, the effective
Lagrangian for the boson fields becomes
\begin{eqnarray}
L(\tau) &=& \sum_i \{ r_i(\tau) ( \partial_{\tau}  + i \lambda(\tau) ) r_i(\tau)
- i \lambda_i(\tau) {N \over 2} \}
\nonumber \\
&-& N Tr \ln[ \{\partial_{\tau} - \mu + i \lambda_i(\tau) +
{1\over N} \sum_l V_{il} ( {N \over 2}- r_l(\tau) r_l(\tau) ) \} \delta_{ij} -
{1 \over N} r_i(\tau) T_{ij} r_j(\tau) ]
\label{radial}
\end{eqnarray}

\subsection{Mean-field solution}

The mean-field solution of the model is obtained by assuming that
the boson fields are spatially homogeneous and time-independent:
$r_i(\tau)=b$ and $i \lambda_i(\tau)=\lambda $.
The resulting 
free energy ($F=-k_B T \ln Z$) is
\begin{equation}
F^{MF}(b,\lambda)=
-{N \over \beta}  \sum_{{\bf k}, \omega_n}
 \ln( \epsilon_{\bf k} - i\omega_n)
 + \lambda (b^2-{N \over 2} )
 \label{MF}
\end{equation}
where $\omega_n$ is a fermion Matsubara frequency. The 
mean-field eigenenergies are
\begin{equation}
\epsilon_{\bf k} = {- t b^2 \over N} T_{\bf k} + \lambda - \mu +
4 V {n \over N}
\end{equation}
with $T_{\bf k}=2( \cos(k_x) + \cos(k_y)  + {t' \over
t} \cos(k_x + k_y) )$ being the Fourier transform of $T_{ij}$ in
units of the nearest-neighbour hopping $t$.

Minimization
of the free energy (\ref{MF}) with respect to
$b$ and $\lambda$ gives 
\begin{equation}
 b^2=N/2-n,\ \ \ \ \ \  \lambda = \sum_{\bf k} f(\epsilon_{\bf
k}) (t T_{\bf k} + 4 V).
\label{bmf}
\end{equation}
 $\mu$ is adjusted to give the
correct electron filling, $n=N \sum_{\bf k} f(\epsilon_{\bf k})$ 
 where $f(\epsilon)$ is the
Fermi-Dirac distribution function.

The mean-field solution describes a renormalized Fermi
liquid. The renormalization of the band is controlled by
$b^2$, and the band is shifted from its bare position
by $\lambda$. The overall effect of
the nearest-neighbours Coulomb interaction, $V$, at the mean-field
level, reduces to a constant shift in the chemical potential.
In the case of a quarter-filled band ($n=1/2$) the bandwidth is
reduced to half its bare value.
The  effective mass measured in magnetic oscillation
experiments will then be 
increased by a factor of $m^*/m=1/b^2=2$.
Note that this is much smaller than the effective
mass enhancements that occur in materials described
by the Hubbard model on the anisotropic triangular lattice
at half-filling.\cite{merino}
Therefore, for $N \rightarrow \infty$
the quarter-filled $t-V$ model behaves as a Fermi liquid
with effective masses that are twice the bare ones.
In the next subsection we consider the effect of the leading
1/N corrections.

\subsection{Fluctuations about the mean-field solution}

We now consider how as $V/t$ is increased, the Fermi liquid phase
resulting from the mean-field solution                              
becomes unstable to charge ordering.
 The analysis is similar 
to the treatment of instabilities in the doped
Hubbard model by Tandon {\it et al.} \cite{Tandon}.
 We write the boson
fields in terms of the static mean-field solution, $(b,\lambda)$,
and the dynamic fluctuating parts: $ r_i(\tau)=b + b \delta r_i
(\tau)$, and $i \lambda_i (\tau)= \lambda + i \delta \lambda_i
(\tau)$.
Physically, $\delta r_i(\tau)$ is
related to local fluctuations in the charge density.
 We substitute these expressions in (\ref{radial}),
introducing the Fourier transforms of  $\delta r_i (\tau)$ and
$\lambda_i (\tau)$, and, expanding to second order in the boson
fields, we obtain the effective action      
\begin{equation}
S = F^{MF} + S^{(2)}
\nonumber \\
\end{equation}
where the part of the action due to fluctuations
in the boson fields is
\begin{equation}
S^{(2)} = {1 \over 2 \beta } \sum_{{\bf q}, \nu_n} \left(
\begin{array}{cc} \delta r(-{\bf q}, -\nu_n) &  \delta
\lambda(-{\bf q}, -\nu_n)
\\ \end{array}
\right) \left( \begin{array}{cc}
\Gamma_{rr} & \Gamma_{r\lambda} \\
\Gamma_{\lambda r} & \Gamma_{\lambda \lambda} \\ \end{array}\right)
\left( \begin{array}{cc}
\delta r({\bf q}, \nu_n)  \\
\delta \lambda({\bf q}, \nu_n) \\ \end{array}\right)
\label{action}
\end{equation}
where $\nu_n$ is a boson Matsubara frequency. The elements of
the $\hat{\Gamma}({\bf q}, \nu_n)$ matrix are
\begin{eqnarray}
{\Gamma_{rr} ({\bf q}, \nu_n)} &=&
N[ {2 b^2 \lambda \over N} -{2 b^2 t \over N} \sum_{\bf k} (T_{\bf k-q}
+ {V \over t} V_{\bf k} ) f(\epsilon_{\bf k} )+
\sum_{\bf k}{ f(\epsilon_{\bf k +q}) - f(\epsilon_{\bf k} )
\over \epsilon_{\bf k + q} - \epsilon_{\bf k} - i \nu_n } ( { b^2 t \over N} (
T_{\bf k} + T_{\bf k + q} ) + {2 b^2 V \over N} V_{\bf q} )^2 ]
\nonumber \\
{\Gamma_{r\lambda} ({\bf q}, \nu_n) } &=&
{\Gamma_{\lambda r} ({\bf q}, \nu_n) } = N[ {i 2 b^2 \over N}
+ i \sum_{\bf k} {f(\epsilon_{\bf k +q}) - f(\epsilon_{\bf k})
\over \epsilon_{\bf k + q} - \epsilon_{\bf k} - i \nu_n } ({-t
b^2 \over N} ( T_{\bf k} + T_{\bf k +q} ) - {2 b^2 V \over N}
V_{\bf q} ) ]
\nonumber \\
{\Gamma_{\lambda \lambda} ({\bf q}, \nu_n) } & = &
- N \sum_{\bf k}{ f(\epsilon_{\bf k +q}) - f(\epsilon_{\bf k} )
\over \epsilon_{\bf k + q} - \epsilon_{\bf k} - i \nu_n }
\end{eqnarray}
where $V_{\bf k}= 2 (\cos(k_x)+ \cos(k_y)) $ is the
Fourier transform of $V_{ij}$.
Note that $\Gamma_{\lambda \lambda}$ is the Lindhard
function describing density-density fluctuations
in the renormalised band structure.
Since $b^2$ is of order N (compare equation (\ref{bmf})),
the above expressions show
explicitly how the propagators of the boson fields
 $\hat{D}({\bf q}, \nu_n )=
\hat{\Gamma}^{-1} ({\bf q}, \nu_n)$ are of order O($1/N$), as
they should.

\subsection{Charge ordering instability}

The condition for the stability of the Fermi liquid phase is
that the quadratic form (\ref{action}) is always positive.
Then fluctuations in the charge density will increase the free energy.
Since $ \Gamma_{\lambda \lambda} > 0  $
this is ensured if $\det \hat{\Gamma}({\bf q}, \nu) > 0$
for all wavevectors ${\bf q}$ and frequency $\nu$.
 In order to find
the critical value of $V/t$, which we shall
denote $(V/t)_c$, at which the system becomes unstable
towards static charge ordering, we wish to find
a ${\bf q}$ for which at some value of $V/t$, $\det \Gamma =
\Gamma_{rr} \Gamma_{\lambda \lambda} - \Gamma_{\lambda r}
\Gamma_{r \lambda}=0$, at $\nu =0$.
This condition reduces to:
\begin{eqnarray}
&[&{b^2 t \over N} \sum_{\bf k} f(\epsilon_{\bf k})( T_{\bf k -q} - T_{\bf k}
+ {(V/t)_c} V_{\bf k} ) + { 4 t (V/t)_c b^4 \over N^2} (1- V_{\bf q}) -
{2 t (V/t)_c b^2 \over N}]
\sum_{\bf k} { f(\epsilon_{\bf k +q}) - f(\epsilon_{\bf k}) \over
\epsilon_{\bf k +q} - \epsilon_{\bf k} }
\nonumber \\
&-&{2 b^4 t\over N^2} \sum_{\bf k} { f(\epsilon_{\bf k +q})
- f(\epsilon_{\bf k}) \over \epsilon_{\bf k +q} - \epsilon_{\bf k} }
(T_{\bf k} + T_{\bf k + q}) + 2 {b^4 \over N^2}= 0
\label{det}
\end{eqnarray}
where $\lambda$ and $b^2$ are the solution to the mean-field equations.

We now concentrate on the case ${\bf q} =(\pi, \pi)$,
which is relevant to the formation of a charge ordered state
at quarter filling in the $\theta$-type materials (see Fig. \ref{fig3}).
For the square lattice case ($t'=0$),
 $\epsilon_{\bf k + q}=-\epsilon_{\bf k}$,
and Eqn. (\ref{det}) reduces to
\begin{equation}
[(1-{(V/t)_c \over 2}) \int_{-b^2 {W \over 2}}^{b^2 {W \over 2}}
 d \epsilon \rho_{\sigma}
 (\epsilon) \epsilon f(\epsilon) + t (V/t)_c 
 (10 {b^4 \over N^2} - {b^2 \over N}) ]
\int_{-b^2 {W \over 2}}^{b^2 {W \over 2}} d \epsilon {\rho_{\sigma}(\epsilon)
f(\epsilon) \over \epsilon }
+ {b^4 \over N^2} =0
\label{square}
\end{equation}
where $\rho_{\sigma}(\epsilon)$ is the density of states at the
Fermi surface per spin channel of the renormalized metal and
$W$ is the bare bandwidth of the metal.

Before solving this equation numerically some
insight can be gained by considering the
case of a constant density of states.
Taking a density of states of the form
$\rho_{\sigma}(\epsilon)
={1 \over  b^2 W }$, if $-b^2 W/2 \le \epsilon \le b^2 W/2$
and 0 otherwise, eq.(\ref{square}) can be simplified further.  
For this case, the critical ratio $(V/t)_c$, at
which $(\pi,\pi)$ charge ordering occurs for a given
electron band filling $n$ is
\begin{equation}
(V/t)_c
=  { {-4 ({N \over 2} -n)^2 \over N} - 2 n ( {n \over
N} -1) \ln(1-2 {n \over N})  \over [ n ( 1- {n \over N}) + { 10
({N \over 2}-n) \over N} -1] \ln(1-2 {n \over N})  }
\label{Vcrit}
\end{equation}

For $N=2$, $(V/t)_c$ diverges at $n=0$, $n=1$ and $n=0.899$.
While the divergence at $n=0$ appears because
it is not possible to have charge ordering when there
is no charge in the lattice, the divergence at $n=1$
is a consequence of the condition that there can be, at most, one
electron at each lattice site: a charge ordered state
of alternating singly and doubly occupied sites would cost
infinite energy to be formed.
The divergence at $n=0.899$ is non-trivial and presumably is a consequence
of the finding made by Tandon {\it et al.}\cite{Tandon}
that close to half-filling, $n \ge
0.88$, the $1/N$ fluctuations drive the $U \rightarrow \infty$ Hubbard
model into phase separation. Hence, the creation of a charge ordered state
is forbidden by the breakdown of periodicity in the charge
distribution of the lattice.
At a quarter filling ($n=1/2$) equation (\ref{Vcrit}) gives
$(V/t)_c = 0.78$.
This can be compared  with the value of $(V/t)_c = 0.69$
obtained from solving (\ref{square}), with the actual density of
states for the square lattice.

\subsection{Effect of the diagonal hopping}

We can include the effect of a next-nearest neighbour hopping
interaction in the analysis presented above. We have solved
equation (\ref{det}), for different values of the $t'/t$ ratio.
Figure \ref{fig5} 
shows how the critical value, $(V/t)_c$, increases
as the ratio, $t'/t$ is increased. The critical value changes
from 0.68 for the square lattice to $(V/t)_c \approx 0.95$ for the
triangular lattice ($t'/t=1$). This can be understood if we plot
the Fermi surfaces for different ratios of the hopping integrals
for fixed band filling: $n=1/2$ (see Fig. \ref{fig6}). While for
the square lattice case, $t'/t=0$, the Fermi surface shows some
remnants of the perfect nesting property present at half-filling,
it gradually elongates along the $k_y=-k_x$ direction becoming
elliptical as $t'/t$ is increased.\cite{aside} This effect makes it even
harder to connect two points at the Fermi surface by the
wavevector ${\bf q}=(\pi, \pi)$. The phase diagram in
Fig. \ref{fig5}, also shows that it is possible to go from the
metallic phase to the charge ordered state by varying either the
$t'/t$ or $V/t$ ratios. 
Because the dependence of
$(V/t)_c $
on $t'/t$ is weak we conclude that the ratio $V/t$ plays a more important
role than $t'/t$ in driving the metal-insulator
transition, within the large-N approach used here.
Hence, as a first approximation we are justified
in neglecting the effect of the diagonal hopping,
as was done in Section \ref{afm}.

Mori\cite{moriV} found that the Coulomb repulsion $V$ varied little
with the angle $\theta$. This is because it scales
roughly with the inverse of the distance between
the molecules. In contrast the hopping integrals
$t_p$ and $t_c$ depend strongly on $\theta$.
$t_p$ varies by about a factor of five as $\theta$
varies from 100 to 140 degrees. This is because 
the overlap integral depends exponentially on the distance
between the molecules. Hence, the main effect of varying $\theta$
is to change the band width. This is what will be driving the metal-insulator
transition.

For $\theta$-(BEDT-TTF)$_2$Cu$_2$(CN)[N(CN)$_2$]$_2$
the measured charge gap\cite{komatsu} is about 200 meV,
suggesting that $V$ is of the order of 100 meV.
Assuming that $V$ does not vary much between materials
the above calculations suggest that the critical value
of the hopping integral $t$ is about 100 meV.
This is consistent\cite{caution} with the values in Table \ref{table1};
i.e., it is quite possible that the materials listed 
there are close to the metal-insulator transition,
as is observed experimentally.

\section{ Relevance to
 $\beta''$-(BEDT-TTF)$_2$SF$_5$YSO$_3$}

 The family $\beta''$-(BEDT-TTF)$_2$SF$_5$YSO$_3$
has been studied\cite{ward} with Y = CH$_2$CF$_2$, CH$_2$, CHF.
The first material 
 is superconducting with a transition temperature of 5.2 K.
Y = CH$_2$ is insulating with a charge
gap of 56 meV and evidence for charge ordering is
found in the fact that alternate molecules
have a different  bond length
and phonon frequency associated with the
central carbon double bond.\cite{charge} The charges are estimated
to be $+0.6e$ and $+0.4e$ where $e$ is the electronic
charge.   
Below room temperature the spin susceptibility decreases 
monotonically, consistent with a spin gap of 8 meV.
Y=CHF is a bad metal and may undergo a metal-semiconductor
transition below 10 K. It is estimated that 
alternate molecules have charges of $+0.47e$ and $+0.53e.$
A recent experimental study\cite{schlueter}
 estimated charges of $+0.43e$ and $+0.57e$
in Y = CH$_2$CF$_2$.
The Fermi surface of the metallic phase 
of Y = CH$_2$CF$_2$ has been mapped out
using angular-dependent magnetoresistance and magnetic
 oscillations.\cite{wosnitza5}
However, the metallic phase differs significantly
from a conventional Fermi liquid.
First, in contrast to most BEDT-TTF metals,\cite{mck}
even at a temperature as low as 14 K
no Drude peak is present in optical conductivity.\cite{dong}
Second, anomalous properties of the magnetoresistance
were recently interpreted in terms of a
magnetic field induced superconductor-insulator 
transition.\cite{wosnitza2}
The temperature dependence of the penetration depth
in the superconducting phase 
was recently found to go like $T^3$ at low
temperatures.\cite{prozorov} This is inconsistent
with an s-wave state, but also deviates significantly
from the linear temperature dependence expected 
for a d-wave state.

The arrangement of the BEDT-TTF molecules within a 
layer of the family $\beta''$-(BEDT-TTF)$_2$SF$_5$YSO$_3$
are shown in Figure \ref{fig4}.  Table \ref{table2}
lists values of the hopping integrals calculated
in the H\"uckel approximation.
Note that generally the diagonal terms a and a' are
smaller than the vertical and horizontal terms.
Hence, as a first approximation the system can be
described by an anisotropic square lattice.
However, we note that the main difference between
the hopping parameters for the three different
anions in Table \ref{table2} is that the diagonal terms
 a and a' vary as the anion  is changed.
This will change 
the proximity to the charge ordering instability
and so lead to the three different ground states
for these materials.

\section{Conclusions }

In summary, we have argued that the essential 
physics of the electronic and magnetic properties
of layered molecular crystals of the $\theta$ type
can be captured by an extended Hubbard model on
the square lattice and at quarter filling.
For large Coulomb repulsion ($V \gg t$) between nearest neighbours,
the ground state is a charge-ordered insulator.
Antiferromagnetic interactions arise due a novel
fourth-order exchange process.
A slave boson treatment was given of the SU(N) generalization
of the model. It was found that for sufficiently small $V/t$
the metallic phase was stable against charge ordering.

We briefly mention the relation of this work to a
recent paper of
Mazumdar, Clay, and Campbell\cite{mazumdar}.
They have studied the
extended Hubbard model at quarter filling on an
anisotropic square lattice
and discuss its relevance to a wide range of organics, but not
those considered here. They argue that in the real
materials the nearest-neighbour Coulomb repulsion 
is smaller than the critical value necessary
to form the charge ordered state considered here.
Coupling to phonons produces an insulating phase 
with a different kind of charge order (a bond-order-wave). 
X-ray scattering experiments which can resolve
the charge on individual molecules (due to different
bond lengths) should be able to distinguish
these two different ground states.
There is controversy about whether nuclear magnetic
resonance measurements can distinguish these two
charge orderings.\cite{mazumdar2}
 The charge distribution
observed\cite{ward} for $\beta''$-(BEDT-TTF)$_2$SF$_5$CH$_2$SO$_3$
is consistent with charge ordering considered here.

\begin{figure}
\centerline{\epsfxsize=9cm \epsfbox{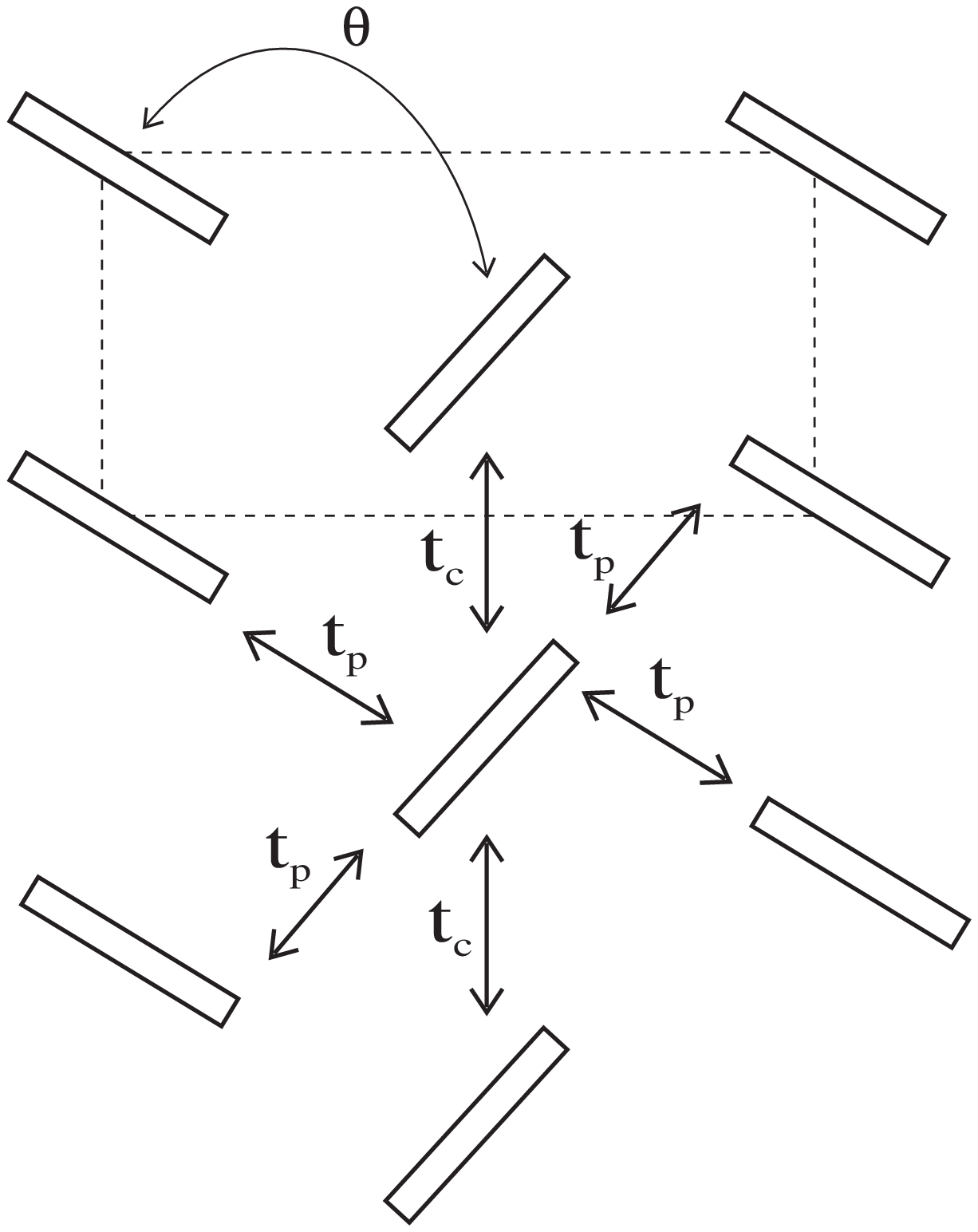}}
\caption{
Arrangement of the donor molecules D (for example BEDT-TTF)
within a layer for the $\theta$-D$_2$X
molecular crystals. 
The dashed rectangle denotes the unit cell.
 Typical values of
the hopping integrals $t_p$ and $t_c$ are given in Table \ref{table1}.
Note that this geometry defines a tight binding model on
an anisotropic triangular lattice which also can
be viewed as a square lattice with hopping along one
of the diagonals.
The angle $\theta$, and consequently the value of $t_p$, varies with pressure
or change in anion X.
}
\label{fig1}
\end{figure}

\begin{figure}
\centerline{\epsfxsize=9cm \epsfbox{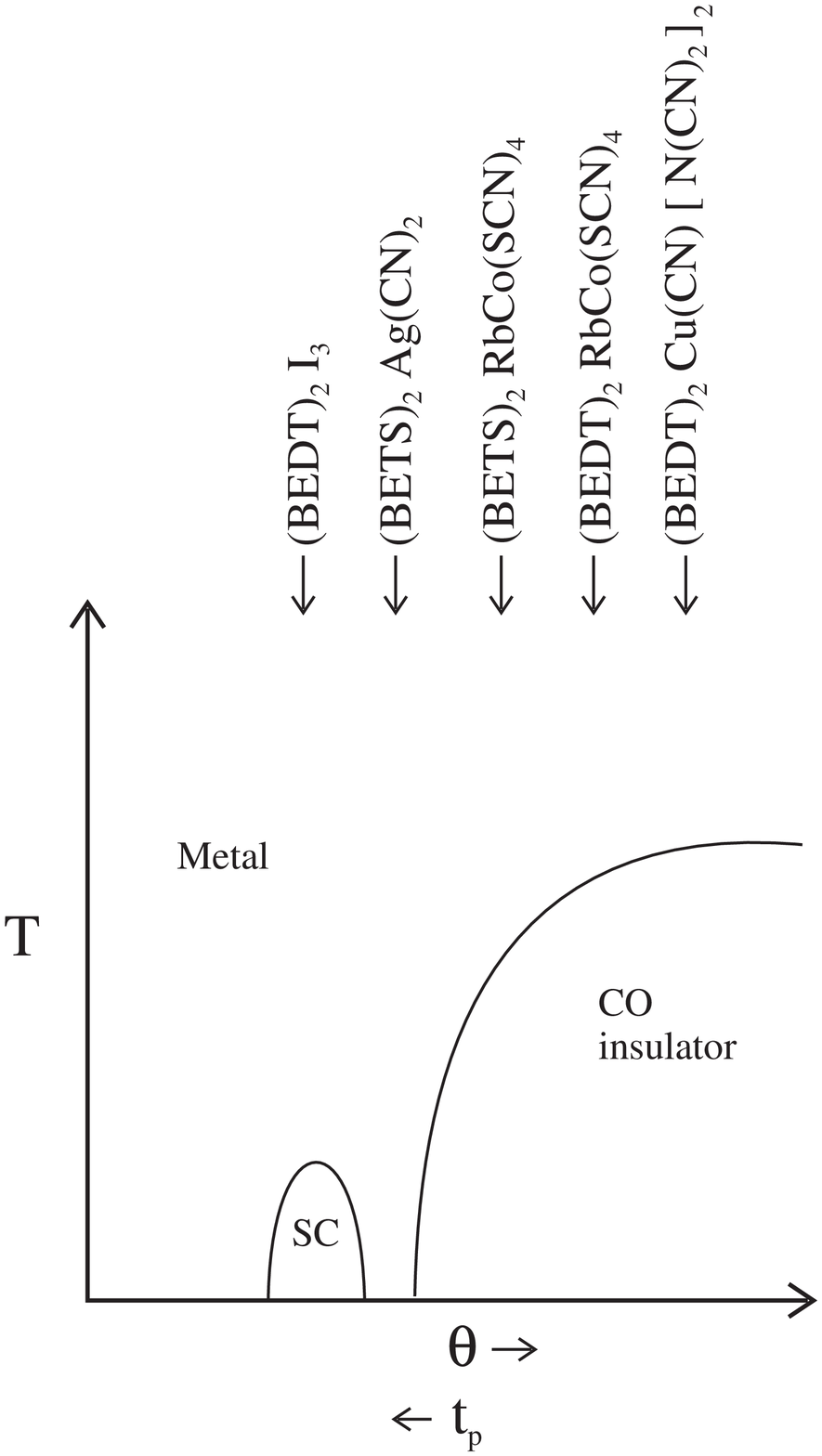}}
\caption{
Schematic phase diagram of the $\theta$-(BEDT-TTF)$_2$X
and $\theta$-(BETS)$_2$X families\cite{mori2,mori4}
showing competition between metallic, superconducting (SC),
and charged ordered (CO) insulating phases.
The horizontal axis is proportional to the angle $\theta$
(see Fig. \protect\ref{fig1}) which is related to the
hopping integral $t_p$.
Generally, increasing $\theta$ decreases the bandwidth and
so increases the importance of
the electronic correlations.
The vertical arrows denote the location of various materials
at ambient pressure.
The effect of pressure is to drive each material 
towards the right.
}
\label{fig2}
\end{figure}

\begin{figure}
\centerline{\epsfxsize=9cm \epsfbox{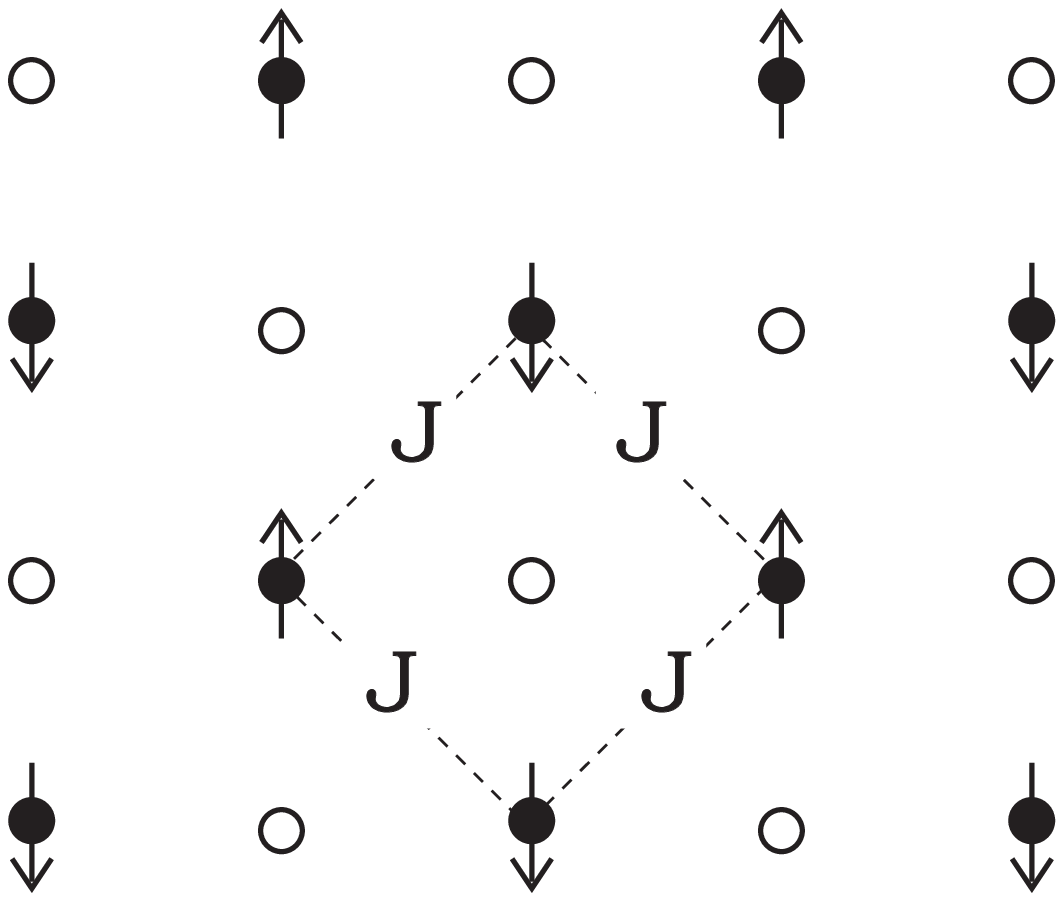}}
\caption{
Charge ordered insulating 
ground state of the extended Hubbard model
at quarter filling in the limit $t \ll V \ll U$.
An antiferromagnetic interaction $J$ occurs between spins
along the diagonals.
The spin degrees of freedom are described by
the antiferromagnetic Heisenberg model on
the square lattice.
}
\label{fig3}
\end{figure}

\begin{figure}
\centerline{\epsfxsize=9cm \epsfbox{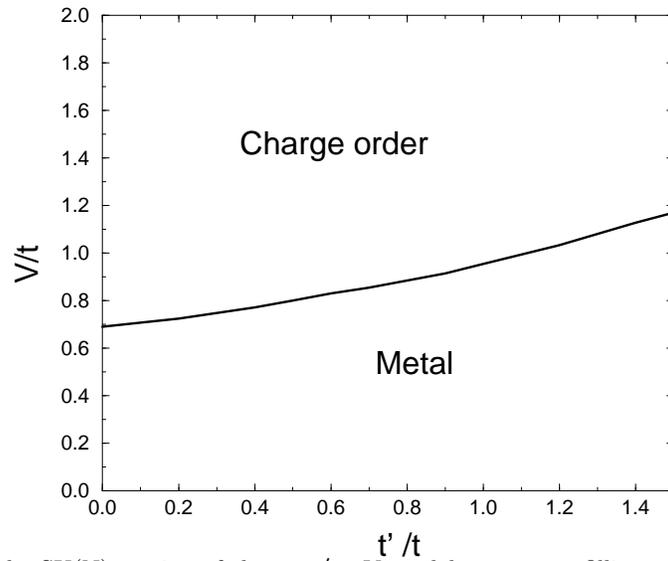}}
\caption{
Phase diagram for the SU(N) version of the 
$t-t'-V$ model at quarter filling and zero temperature,
to leading order in 1/N with N=2.
This shows that the diagonal hopping $t'$ ($t_c$ in
Figure \protect\ref{fig1}) has little effect on
the critical value of $V/t$ at which the
metallic phase becomes unstable to the charge ordering
shown in Figure \protect\ref{fig3}.
}
\label{fig5}
\end{figure}

\begin{figure}
\hskip 0.1cm \psfig{file=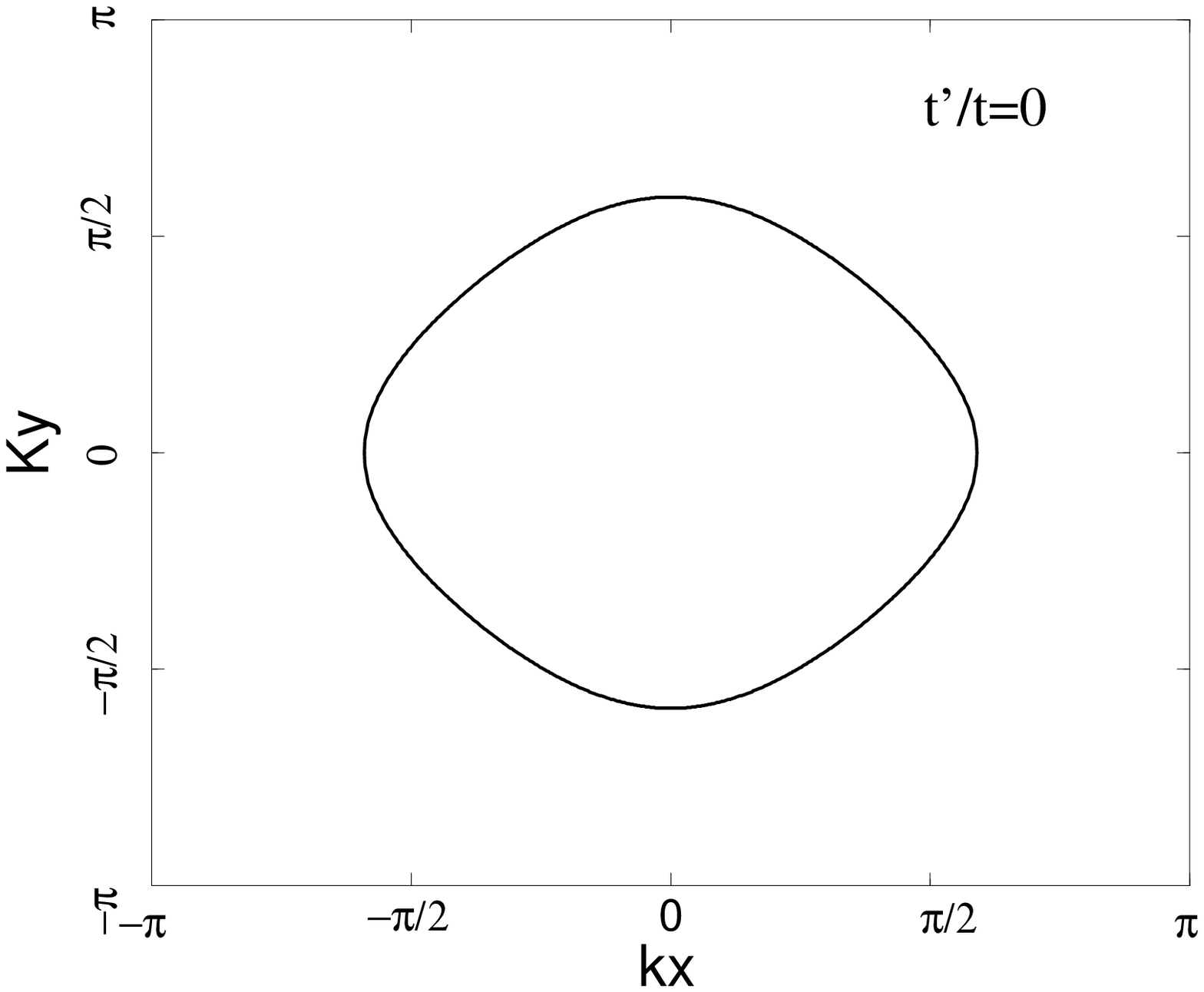,width=5.2cm,height=5cm} \hskip 0.1cm \psfig{file=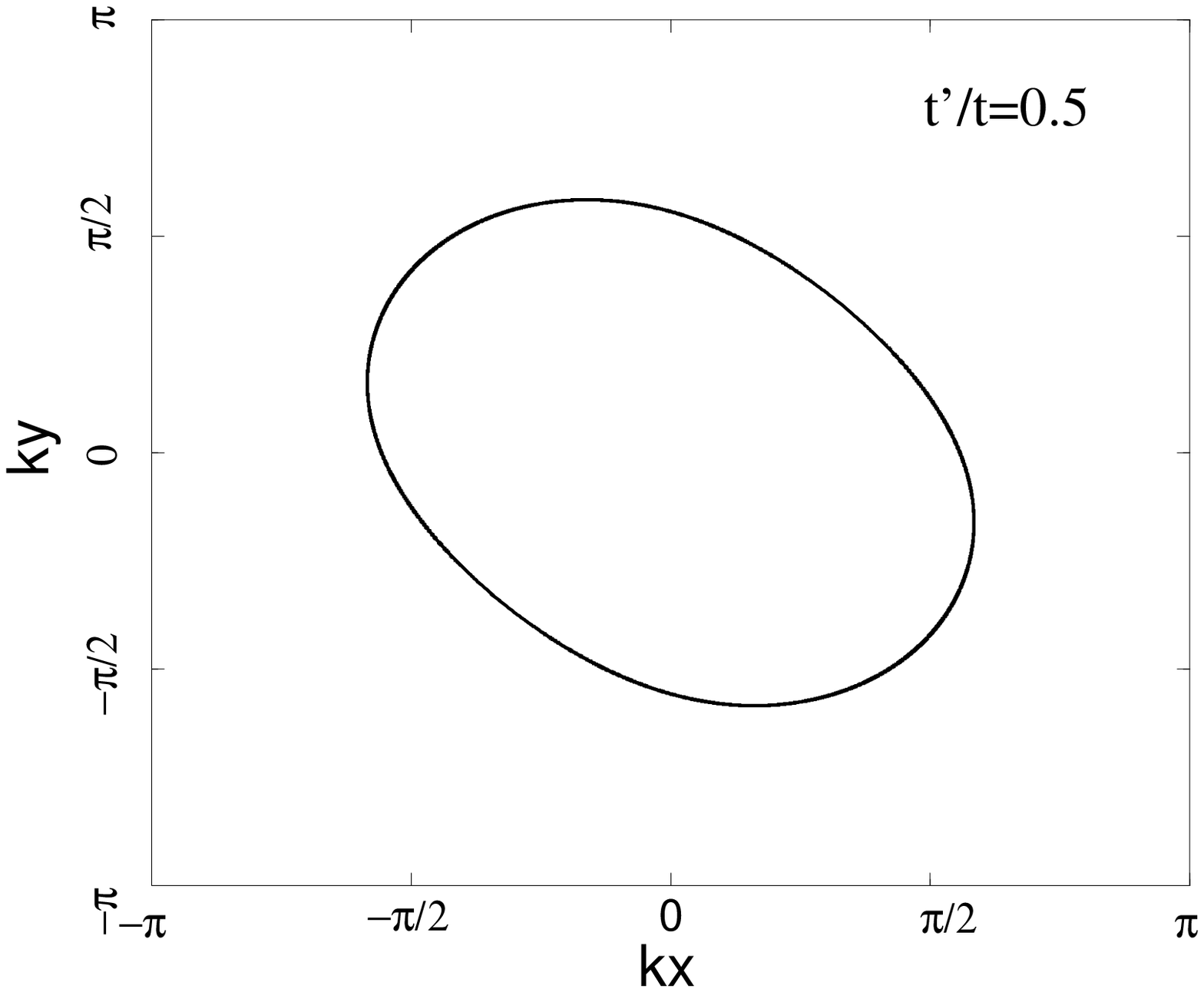,width=5.2cm,height=5cm} \hskip 0.1cm \psfig{file=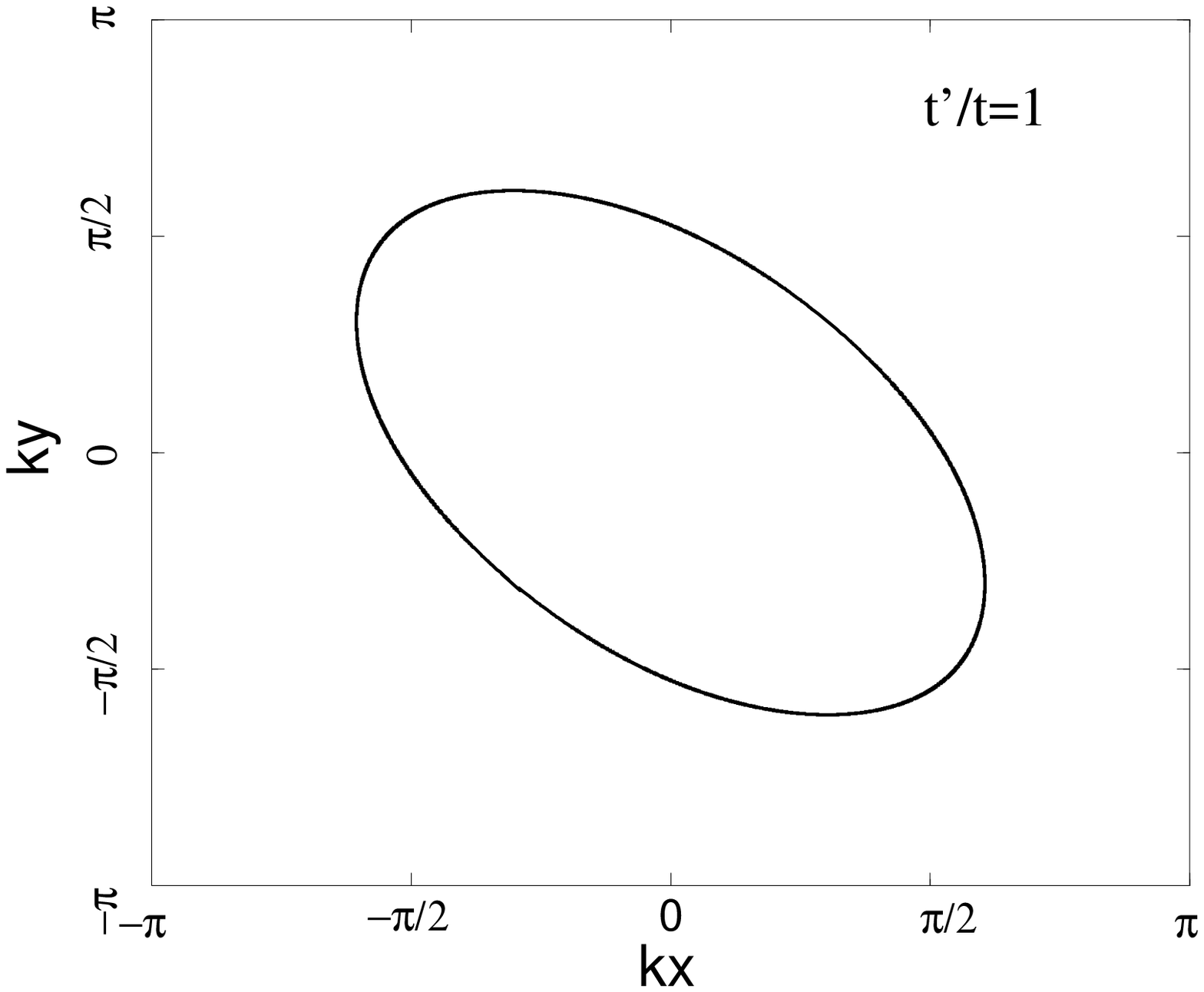,width=5.2cm,height=5cm}
\caption{Evolution of the Fermi surface for an anisotropic
triangular lattice
as the ratio between the next-nearest neighbour and the nearest
neighbours hoppings, $t'/t$, is varied.
The band is kept at quarter-filling for all three cases.
}
\label{fig6}
\end{figure}

\begin{figure}
\centerline{\epsfxsize=9cm \epsfbox{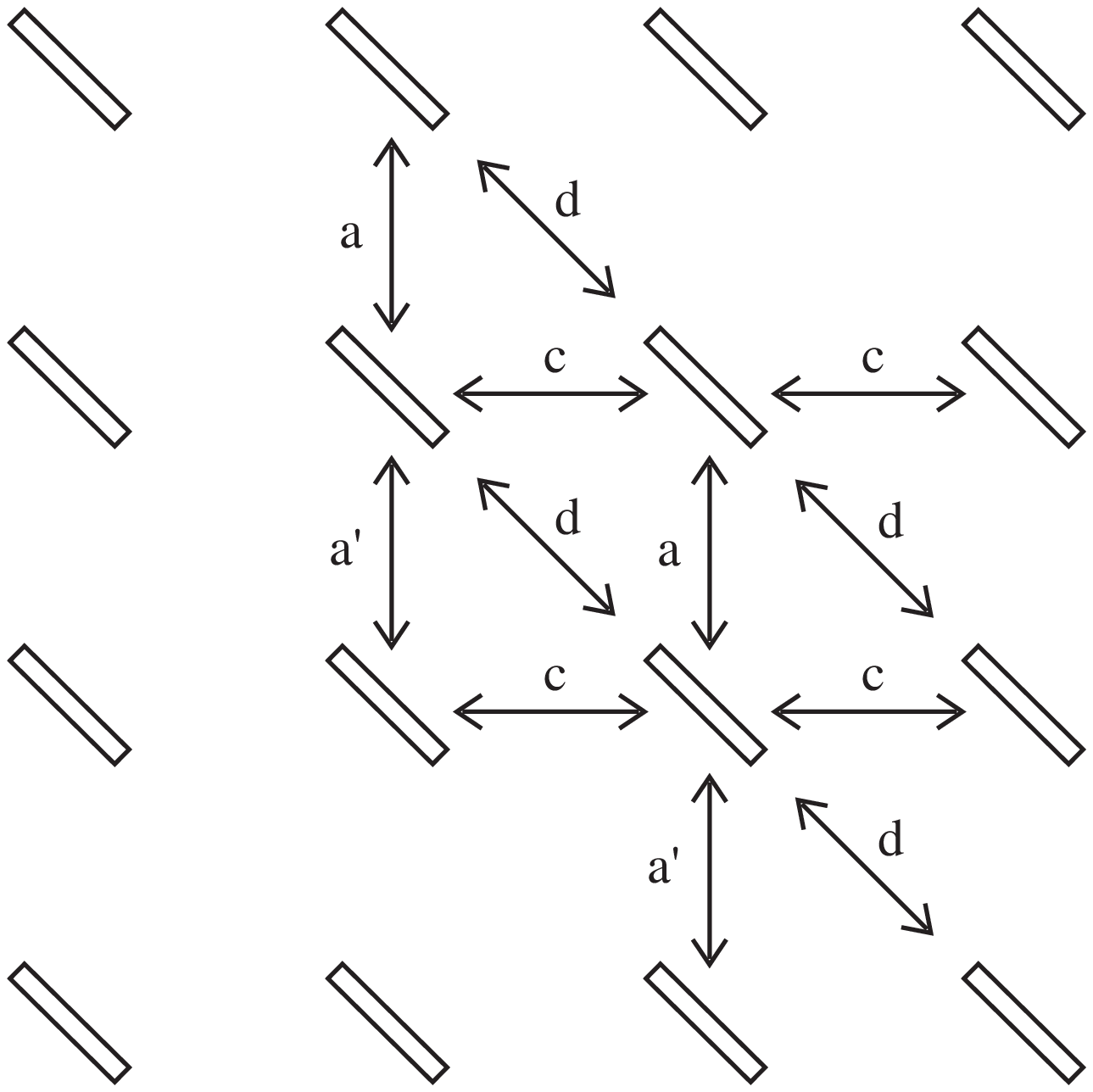}}
\caption{
Arrangement of the BEDT-TTF molecules
within a layer for the 
 $\beta''$-(BEDT-TTF)$_2$SF$_5$YSO$_3$
family of molecular crystals. Typical values of
the hopping integrals are given in Table \ref{table2}.
In some of the materials the unit cell is larger
and so the hopping integrals can have two values.
}
\label{fig4}
\end{figure}

We acknowledge that
the actual $\theta$ type materials are more complicated than the simplest
Hubbard model considered here.
For example, along the diagonals of the square lattice (corresponding
to the vertical direction in Figure \ref{fig1}) there is 
also Coulomb repulsion. In fact, Mori\cite{moriV} finds
the corresponding $V$ to be larger than along the horizontal
and vertical directions. Seo\cite{seo} has shown how the
latter can lead to competition between
different charge ordered states (i.e., those associated
with wave vector $(\pi,\pi)$ and
$(0,\pi)$). Also, x-ray scattering suggests
that in 
(BEDT-TTF)$_2$RbM(SCN)$_4$ [M=Co,Zn]
there is a structural transition associated with
the charge ordering and that this changes the electronic 
structure in the insulating phase.\cite{Mori}
However, our view is that the $t-V$ model
on the square lattice captures the essential physics and 
first we need to understand it.

Three outstanding questions concerning the
$t-V$ model at quarter filling need to be answered.

(i) Is there superconductivity in the model?
The idea that proximity to a quantum critical point 
increases the tendency towards
superconducting instabilities is
supported by recent experiments on heavy 
fermion materials.\cite{mathur}
It was first shown by Scalapino, Loh, and Hirsch\cite{scalapino}
that proximity to a spin-density wave or charge-density
wave transition can lead to d-wave superconductivity.
In a future publication we will investigate whether
charge fluctuations near the charge ordering 
transition can produce superconductivity.\cite{bang,perali,azami}

(ii) Are charge ordering, the charge gap, and antiferromagnetism
destroyed at the same critical value of $V/t$?
 In Section \ref{afm} it was shown that for large $V/t$
the ground state has a charge gap, charge ordering,
and antiferromagnetism. In  Section \ref{slave}
it was shown that, for $V/t$ less than a critical
value, the metallic phase is stable, at
least in the large N limit.
It is quite possible that the above three properties
disappear at different values of $V/t$. 
For the case of the quarter-filled extended Hubbard model on a
ladder, numerical calculations found that the charge ordering disappeared
below a non-zero value of $V/t$, but the charge gap did not.\cite{votja} 
This unusual result may be an artefact of the one-dimensionality
of the ladder.
For the square lattice, this issue will probably
be only resolved by careful numerical work.

(iii) Does non-Fermi liquid behavior occur in
the metallic phase near the quantum critical point?
This is generally expected\cite{sachdev0} and is observed
in heavy fermion materials.\cite{coleman}
Slave boson theory has been used to show that in the doped Hubbard model
near a charge instability the quasi-particle scattering becomes
singular leading to anomalous metallic properties.\cite{castellani}

\acknowledgements

We thank M. Brunner, R. J. Bursill, O. Cepas,
 C. J. Hamer, C. J. Kepert, R. R. P. Singh
and J. A. Schlueter  for helpful discussions.
This work was supported by the Australian
Research Council.
J. B. Marston was supported by 
NSF grant DMR-9712391 and the
Gordon Godfrey Bequest for Theoretical
Physics at the University of New South Wales.
Most of the figures were produced by Perez Moses.

\newpage

\newpage

\begin{table}
\caption{
Hopping integrals for various $\theta$-type crystals   
calculated by the H\"uckel method.
Two values are given for the case where the unit cell
is larger. The temperature of the metal-insulator transition,
$T_{MI}$, is also given.
Note the general trend, observed by Mori\protect\cite{Mori}, that as 
$t_p$ increases $T_{MI}$ decreases.\protect\cite{careful}
}
\vskip0.1in
\label{table1}
\begin{tabular}{ldddd}
Material & $t_p$ (meV) &	 $t_c$ (meV) & Reference & $T_{MI} $ (K)\\
\hline
(BEDT-TTF)$_2$I$_3$	&  42& 64 & \onlinecite{kobayashi} & -- \\
(BETS)$_2$Ag(CN)$_2$ & 392,398 & -1,38 & \protect\onlinecite{mori4}& $< 4$ \\
(BETS)$_4$Cu$_2$Cl$_6$ & 380-467  & 
-12 - +56 & \protect\onlinecite{kobayashi3}& $< 4$ \\
(BETS)$_2$CsCo(SCN)$_4$	& 366 & -2  & \protect\onlinecite{mori4}& 10? \\
(BETS)$_2$CsZn(SCN)$_4$	& 372 & -10 & \protect\onlinecite{mori4} & 10? \\
(BETS)$_2$RbCo(SCN)$_4$	& 382 & -72 & \protect\onlinecite{mori4} & 20 \\
(BETS)$_2$RbZn(SCN)$_4$	& 347 & -46 & \protect\onlinecite{mori4} &  ? \\
(BEDT-TTF)$_2$CsCo(SCN)$_4$	& 106 & -5 & \onlinecite{Mori} & 20 \\
(BEDT-TTF)$_2$CsZn(SCN)$_4$	& 108 & -10 & \onlinecite{Mori}& 20  \\
(BETS)$_4$TaF$_6$ &  -30 & 110 & \protect\onlinecite{kobayashi2}& 70 \\
(BEDT-TTF)$_2$RbCo(SCN)$_4$	& 99  & -33 & \onlinecite{Mori} & 190 \\
(BEDT-TTF)$_2$RbZn(SCN)$_4$	& 94  & -24 & \onlinecite{Mori} & 190 \\
(BEDT-TTF)$_2$Cu$_2$(CN)[N(CN)$_2$]$_2$	& 79 & -30
  & \onlinecite{komatsu} & 220 \\
(BEDT-TTF)$_2$TlCo(SCN)$_4$	& 100 & -48 & \onlinecite{Mori} & 250 \\
(BDT-TTP)$_2$Cu(NCS)$_2$ & -86,-91 & -41 & \onlinecite{yamaura} & 250 \\ 
(C$_1$TET-TTF)$_2$Br & -54,-43 & -58 & \onlinecite{yamaura} & $> 300$ \\
\end{tabular}
\end{table}

\begin{table}
\caption{
Hopping integrals in Figure \protect\ref{fig4} for 
 the family $\beta''$-(BEDT-TTF)$_2$SF$_5$YSO$_3$
calculated by the extended H\"uckel method
in Reference \protect\onlinecite{ward}.
At low temperatures the
materials are a superconductor, a bad metal,
and a charge ordered insulator, respectively.
}
\vskip0.1in
\label{table2}
\begin{tabular}{ldddd}
Y & c (meV) &	 d (meV) &    a (meV) &    a' (meV)\\
\hline
CH$_2$CF$_2$ & 260 & 140 & 120 & 55 \\
CHF & 260 & 130 & 35, 86 & 95, 100\\
CH$_2$ & 260 & 120 & 85  & 12 \\
\end{tabular}
\end{table}


\begin{references}
\bibitem{sulphur}
If all the sulfur atoms in the BEDT-TTF molecule
are replaced with selenium one obtains
the BETS molecule.
\bibitem{caution}
One should be cautious about taking the H\"uckel and
extended H\"uckel values as definitive. Based on
comparison with results from the local density
approximation it is argued in Reference
\protect\onlinecite{merino}
that the H\"uckel methods tend to systematically
underestimate the hopping integrals.
\bibitem{aside} As an aside, we note that when one produces 
a small dimerisation along the diagonal direction, a folding
of the Fermi surface occurs, producing an open sheet and
closed pocket, consistent with the $\alpha$ and $\beta$ orbits
seen in magnetic oscillation experiments.\protect\cite{tamura}
\end{references}
\end{document}